\documentclass{aa}

\usepackage{txfonts}
\usepackage[dvips]{graphicx}
\usepackage{natbib}
\bibpunct{(}{)}{;}{a}{}{,} 

\begin{document}

\title{Mid-infrared imaging of 25 local AGN with VLT-VISIR
       \thanks{Based on ESO observing programmes 075.B-0844(C) and 
               077.B-0137(A)}}
\author{Hannes Horst \inst{1,2,3,4}
  \and Wolfgang J. Duschl \inst{1,5}
  \and Poshak Gandhi \inst{6}
  \and Alain Smette \inst{4}}
\authorrunning{H. Horst et al.}
\offprints{H. Horst, \email{hhorst@astrophysik.uni-kiel.de}}
\institute{Institut f\"ur Theoretische Physik und Astrophysik, 
       Christian-Albrechts-Universit\"at zu Kiel, Leibnizstr. 15, 24098 Kiel,
       Germany
  \and Zentrum f\"ur Astronomie, ITA, Universit\"at Heidelberg, 
       Albert-Ueberle-Str. 2, 69120 Heidelberg, Germany
  \and Max-Planck-Institut f\"ur Radioastronomie, Auf dem H\"ugel 69, 53121
       Bonn, Germany
  \and European Southern Observatory, Casilla 19001, Santiago 19, Chile
  \and Steward Observatory, The University of Arizona, 933 N. Cherry Ave, 
       Tucson, AZ 85721, USA
  \and RIKEN Cosmic Radiation Lab, 2-1 Hirosawa, Wakoshi Saitama 351-0198,
       Japan}
\date{Received 00.00.0000 / Accepted 00.00.0000}
\abstract{}
         {High angular resolution N-band imaging is used to discern the torus
          of active galactic nuclei (AGN) from its environment in order to
          allow a comparison of its mid-infrared properties to the expectations
          of the unified scenario for AGN.}
         {We present VLT-VISIR images of 25 low-redshift AGN of different 
          Seyfert types, as well as N-band SEDs of 20 of them. In addition,
          we compare our results for 19 of them to \emph{Spitzer} IRS spectra.}
         {We find that at a resolution of $\sim 0\farcs35$, all the nuclei of 
          our observed sources are point-like, except for 2 objects whose
          extension is likely of instrumental origin. For 3 objects, however, 
          we observed additional extended circumnuclear emission, even though
          our observational strategy was not designed to detect it. 
          Comparison of the VISIR photometry and \emph{Spitzer}
          spectrophotometry indicates that the latter is affected by extended
          emission in at least 7 out of 19 objects and the level of 
          contamination is $(0.20 \sim 0.85) \cdot F_{\mathrm{IRS}}$. 
          In particular, the $10 \, \mu$m silicate emission feature seen in the
          \emph{Spitzer} spectra of 6 type I AGN, possibly 1 type II AGN and 2 
          LINERs, also probably originates not solely in the torus but also in 
          extended regions.}
         {Our results generally agree with the expectations from the unified 
          scenario, while the relative weakness of the silicate feature 
          supports clumpy torus models. Our VISIR data indicate that, for
          low-redshift AGN, a large fraction of \emph{Spitzer} IRS spectra are
          contaminated by extended emission close to the AGN.}


\maketitle

\section{Introduction}

The unification model for active galactic nuclei (AGN) interprets the 
different appearance of Seyfert 1 and Seyfert 2 galaxies uniquely 
as the result of an orientation effect \citep{anton93,barthel94,urry96}. The 
central engine is considered to be surrounded by an optically and geometrically
thick molecular torus that appears as a grey body, mainly emitting in the
infrared; the SED peaks around $20 \, \mu$m in terms of energy emitted per unit
wavelength. Therefore, one way to study the physical properties of the putative
torus, is to observe AGN in the mid infrared (MIR) regime between 
$\sim 8 \, \mu$m and $\sim 30 \, \mu$m. 

In recent years, this field progressed rapidly thanks to the arrival of 
powerful new instruments, mainly the \emph{Spitzer} space telescope on one 
hand and high-resolution imagers and spectrographs at 10m-class telescopes 
(COMICS, LWS/LWIRC, Michelle, T-ReCS, VISIR) as well as the interferometric
instrument MIDI on the other hand. Thus, it has become feasible to compare 
observations in detail to radiative transfer modelling of dusty tori 
\citep[e.g.][]{schartmann05,hoenig06,fritz06} and constrain the geometry of
the torus \citep[e.g.][]{buchanan06,polletta07b,treister08} as well as test the
applicability of the unified scenario in general \citep{haas07}. 

In two recent publications, \citet{horst06,horst08a}, hereafter called Paper~I
and Paper~II, respectively,\defcitealias{horst06}{Paper~I} 
\defcitealias{horst08a}{Paper~II} we investigated the correlation between the
mid-IR and hard X-ray luminosities in AGN and its implications for dusty
torus models. To this end, we observed 25 local ($z < 0.1$) AGN with the
VISIR \citep{lagage04} instrument at the VLT and found a highly significant
correlation between their rest frame $12.3 \, \mu$m and $2-10$ keV
luminosities. Moreover, we found the luminosity ratio $L_{\mathrm{MIR}}$ /
$L_{\mathrm{X}}$ to be independent of Seyfert type and luminosity, in agreement
with the studies by \citet{krabbe01}, \citet{alonso02} and \citet{lutz04}. Our 
sample has recently been expanded to 41 VISIR detections, including many 
Compton-thick AGN, by \citet{gandhi08}.

In the present article, we present the mid-IR properties of this sample of 
objects in more detail. Typically, we achieved a spatial resolution of 
$\sim 0\farcs35$ FWHM, corresponding to a resolved linear size ranging from 
6 pc to 510 pc, depending on the distance of the objects. Since our main goal 
was to derive the luminosity of the central point source, our imaging is not 
very deep. In a few cases, however, we detected and resolved extra-nuclear 
emission.

For 20 of the presented  objects, photometry in different N-band filters is 
available which enables us to reproduce the overall shape of the N-band SEDs of
these objects. When possible, we compare our photometric data to spectra 
obtained with the IRS instrument \citep{houck04} aboard \emph{Spitzer}. While 
in many cases, \emph{Spitzer} and VISIR data are in good agreement, in others 
they are not. We will discuss the possible consequences of such disagreements 
for the study of AGN tori with \emph{Spitzer} data.

Throughout this paper we assume $H_{0} = 73$ km s$^{-1}$ Mpc$^{-1}$, 
$\Omega_{\Lambda} = 0.72$ and $\Omega_{\mathrm{m}} = 0.24$ \citep{spergel06}. 

\section{Observations and data analysis} \label{data}

Two samples of AGN were observed, one between April and August 2005 and the
other one during the same period in 2006. Details on the target selection and 
observing conditions are given in \citetalias{horst06} for the first sample 
and \citetalias{horst08a} for the second sample, respectively. 

We used the standard imaging template of VISIR, with parallel chopping and 
nodding and a chop throw of $8\arcsec$. In order to get the best possible 
angular resolution, the small field objective ($0\farcs075$ / pixel) was used. 
Bright AGN were observed in three filters in order to allow a reconstruction 
of their spectral energy distribution (SED) in the MIR. Due to time 
constraints, faint objects could only be observed in one filter. All 
observations were executed in service mode with required observing conditions
of clear sky and $0\farcs8$ seeing or less. The average airmass was 1.15, with 
no observation being executed at an airmass above 1.3. Science targets and 
photometric standards were all observed within 2 h of each other and with a 
maximum difference in airmass of 0.25. For most observations, however, 
differences in both time and airmass are much smaller than these values. 
Some exposures had to be re-executed as the atmospheric conditions were not
within specified constraints. In these cases, we only present the data 
obtained under the best conditions.

We reduced science and standard star frames using the pipeline written by Eric
Pantin \citep[private communication, also see][]{pantin08} for the VISIR 
consortium. To eliminate glitches, the pipeline applies a bad pixel mask and 
removes detector stripes. Subsequently, we removed background variations using 
a 2-dimensional, 6th degree polynomial fit. For objects observed in unstable 
conditions, we treated each nodding cycle separately as the background pattern 
sometimes changed between two consecutive cycles. The count rate for one full 
exposure was calculated as the mean of all 3 beams from all nodding cycles of 
this exposure. As an error estimate we use the standard deviation of these. In 
order to minimise the effect of residual sky background we chose relatively 
small apertures ($\approx 10$ pixels $= 1 \farcs 27$) for the photometry and 
corrected the obtained count rates using the radial profiles of standard 
stars. Finally, we calibrated our photometry using the same standard stars. The
conversion factor counts/s / Jy proved to be very stable: for each individual 
filter, variations were less than 10 \% rms over the whole observing period.

\section{Results} \label{results}

In Table \ref{fluxtable}, we list the measured fluxes and resolved scales for
each object and filter. The fluxes are reproduced from \citetalias{horst06} and
\citetalias{horst08a}. While in these papers, the resolved scales were computed
for a fixed angular resolution of $0\farcs35$, we here use the actual size of
the measured FWHM of the PSF. The scales are rounded to multiples of 5 pc, 
with the exception of \object{Cen A}; for this object the scale was rounded to 
1 pc. As can be seen in the Table, the resolved scale varies from 6 pc 
(\object{Cen A} in SIV) to 510 pc (\object{PG 2130+099} in NeII). We also 
state each object's Seyfert type according to \citet{veron06}. For 
peculiarities of individual objects, consult \citetalias{horst08a}. 
Henceforth, we will denote Sy types 1, 1.2 and 1.5 as 'type I' AGN, and Sy 
types 1.8, 1.9, 2.0 and 1h as 'type II' AGN.

\begin{table}
  \begin{center}
    \caption{Object names, Sy types, VISIR filters, central wavelengths, fluxes
             and resolved scales for all detected sources. References for
             fluxes are \citetalias{horst06} and \citetalias{horst08a}.} 
    \label{fluxtable}
    \scriptsize{
    \begin{tabular}{lclrcr}
      \hline\hline
      Object & Sy type &Filter & $\lambda_c$ [$\mu$m] & Flux [mJy] & 
      Scale [pc] \\
      \hline
      \object{Fairall 9}       &1.2&SIV      & 10.49 & $256.2 \pm 5.4$  & 310\\
      \object{Fairall 9}       &1.2&NeIIref1 & 12.27 & $329.8 \pm 18.0$ & 340\\
      \object{Fairall 9}       &1.2&NeII     & 12.81 & $305.7 \pm 10.4$ & 340\\
      \object{NGC 526a}        &1.9&SIV      & 10.49 & $198.6 \pm 22.0$ & 115\\
      \object{NGC 526a}        &1.9&NeIIref1 & 12.27 & $275.3 \pm 55.0$ & 135\\
      \object{Mrk 590}         &1.0&SIV      & 10.49 & $75.9  \pm 20.9$ & 120\\
      \object{Mrk 590}         &1.0&PAH2     & 11.25 & $75.0  \pm 2.1$  & 135\\
      \object{Mrk 590}         &1.0&NeII     & 12.81 & $106.3 \pm 13.3$ & 160\\
      \object{NGC 1097}        &L  &NeIIref1 & 12.27 & $28.2  \pm 6.8$  & 30\\
      \object{NGC 3783}        &1.5&SIV      & 10.49 & $568.1 \pm 46.2$ & 65\\
      \object{NGC 3783}        &1.5&PAH2ref2 & 11.88 & $632.2 \pm 21.9$ & 70\\
      \object{NGC 3783}        &1.5&NeIIref1 & 12.27 & $721.8 \pm 67.3$ & 70\\
      \object{NGC 4507}        &1h &SIV      & 10.49 & $523.2 \pm 24.9$ & 75\\
      \object{NGC 4507}        &1h &PAH2     & 11.25 & $589.5 \pm 21.8$ & 75\\
      \object{NGC 4507}        &1h &NeIIref1 & 12.27 & $685.0 \pm 50.1$ & 80\\
      \object{NGC 4579}        &L  &SIV      & 10.49 & $64.2  \pm 11.6$ & 32\\
      \object{NGC 4579}        &L  &PAH2ref2 & 11.88 & $68.5  \pm 13.8$ & 29\\
      \object{NGC 4579}        &L  &NeIIref1 & 12.27 & $60.7  \pm 20.6$ & 30\\
      \object{NGC 4593}        &1.0&SIV      & 10.49 & $331.4 \pm 28.8$ & 60\\
      \object{NGC 4593}        &1.0&PAH2ref2 & 11.88 & $335.4 \pm 26.1$ & 60\\
      \object{NGC 4593}        &1.0&NeIIref1 & 12.27 & $382.4 \pm 73.3$ & 65\\
      \object{NGC 4941}        &2.0&NeIIref1 & 12.27 & $81.3  \pm 6.0$  & 30\\
      \object{IRAS 13197-1627} &1h &SIV      & 10.49 & $527.1 \pm 17.1$ & 120\\
      \object{IRAS 13197-1627} &1h &PAH2     & 11.25 & $674.3 \pm 35.8$ & 125\\
      \object{IRAS 13197-1627} &1h &NeIIref1 & 12.27 & $875.0 \pm 45.8$ & 140\\
      \object{Cen A}           &2.0&SIV      & 10.49 & $642.6 \pm 26.6$ & 6\\
      \object{Cen A}           &2.0&PAH2     & 11.25 & $946.6 \pm 29.2$ & 7\\
      \object{Cen A}           &2.0&NeIIref1 & 12.27 & $1451  \pm 73.1$ & 7\\
      \object{NGC 5135}        &2.0&NeIIref1 & 12.27 & $122.5 \pm 12.2$ & 95\\
      \object{MCG-06-30-015}   &1.5&SIV      & 10.49 & $339.2 \pm 43.7$ & 50\\
      \object{MCG-06-30-015}   &1.5&PAH2     & 11.25 & $392.5 \pm 54.1$ & 50\\
      \object{MCG-06-30-015}   &1.5&NeIIref1 & 12.27 & $392.7 \pm 49.3$ & 55\\
      \object{NGC 5995}        &1.9&SIV      & 10.49 & $296.8 \pm 30.2$ & 180\\
      \object{NGC 5995}        &1.9&PAH2     & 11.25 & $332.9 \pm 47.2$ & 180\\
      \object{NGC 5995}        &1.9&NeII     & 12.81 & $421.1 \pm 60.6$ & 195\\
      \object{ESO 141-G55}     &1.0&SIV      & 10.49 & $160.0 \pm 21.2$ & 255\\
      \object{ESO 141-G55}     &1.0&PAH2     & 11.25 & $169.8 \pm 23.9$ & 245\\
      \object{ESO 141-G55}     &1.0&NeIIref1 & 12.27 & $169.7 \pm 47.1$ & 220\\
      \object{Mrk 509}         &1.5&SIV      & 10.49 & $226.5 \pm 7.7$  & 205\\
      \object{Mrk 509}         &1.5&PAH2     & 11.25 & $235.0 \pm 21.4$ & 210\\
      \object{Mrk 509}         &1.5&NeII     & 12.81 & $269.0 \pm 41.7$ & 245\\
      \object{PKS 2048-57}     &1h &SIV      & 10.49 & $590.6 \pm 19.4$ & 85\\
      \object{PKS 2048-57}     &1h &PAH2     & 11.25 & $752.4 \pm 45.5$ & 85\\
      \object{PKS 2048-57}     &1h &PAH2ref2 & 11.88 & $883.1 \pm 53.0$ & 90\\ 
      \object{PKS 2048-57}     &1h &NeIIref1 & 12.27 & $1040  \pm 63.7$ & 90\\
      \object{PG 2130+099}     &1.5&SIVref1  & 9.82  & $114.6 \pm 19.9$ & 425\\
      \object{PG 2130+099}     &1.5&PAH2     & 11.25 & $173.9 \pm 16.5$ & 475\\
      \object{PG 2130+099}     &1.5&NeII     & 12.81 & $179.1 \pm 30.5$ & 510\\
      \object{NGC 7172}        &2.0&NeIIref1 & 12.27 & $164.9 \pm 27.1$ & 60\\
      \object{NGC 7213}        &L  &SIV      & 10.49 & $283.8 \pm 6.2$  & 35\\
      \object{NGC 7213}        &L  &PAH2     & 11.25 & $264.0 \pm 38.5$ & 40\\
      \object{NGC 7213}        &L  &NeIIref1 & 12.27 & $271.0 \pm 26.5$ & 35\\
      \object{3C 445}          &1.5&SIV      & 10.49 & $168.4 \pm 6.7$  & 325\\
      \object{3C 445}          &1.5&PAH2     & 11.25 & $184.6 \pm 10.4$ & 335\\
      \object{3C 445}          &1.5&NeII     & 12.81 & $205.8 \pm 27.8$ & 385\\
      \object{NGC 7314}        &1h&SIV      & 10.49 & $74.9  \pm 29.4$ & 35\\
      \object{NGC 7314}        &1h&PAH2     & 11.25 & $74.5  \pm 22.0$ & 35\\
      \object{NGC 7469}        &1.5&SIV      & 10.49 & $460.0 \pm 20.0$ & 120\\
      \object{NGC 7469}        &1.5&PAH2     & 11.25 & $487.3 \pm 38.6$ & 130\\
      \object{NGC 7469}        &1.5&NeIIref1 & 12.27 & $626.9 \pm 34.7$ & 115\\
      \object{NGC 7674}        &1h &NeII     & 12.81 & $506.3 \pm 29.4$ & 245\\
      \object{NGC 7679}        &1.9&SIV      & 10.49 & $42.4  \pm 13.0$ & 90\\
      \object{NGC 7679}        &1.9&PAH2     & 11.25 & $43.3  \pm 6.6$  & 100\\
      \object{NGC 7679}        &1.9&NeIIref1 & 12.27 & $45.6  \pm 18.3$ & 135\\
      \hline
    \end{tabular}}
  \end{center}
\end{table}

\subsection{Mid-IR properties}  \label{basics}

\begin{figure*}
  \begin{center}
    \includegraphics[height=24cm]{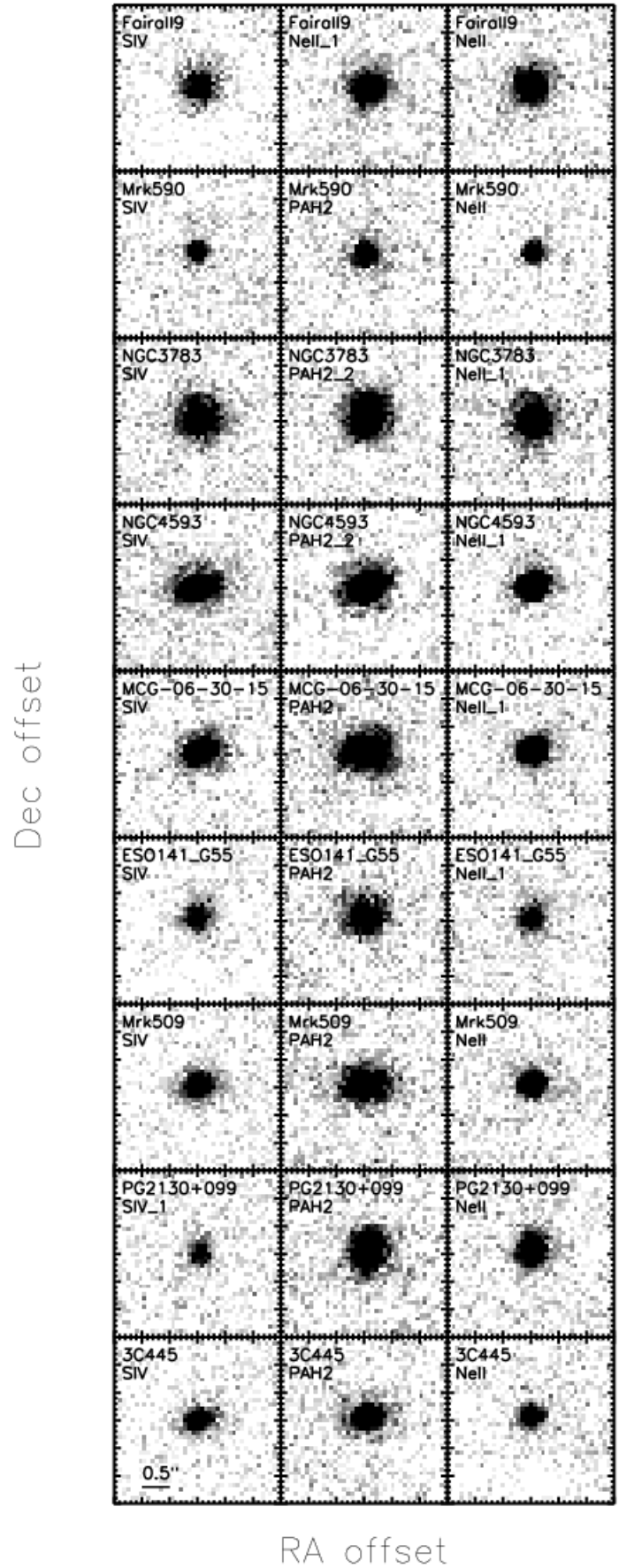}
    \caption[VISIR images of type I AGN]
            {$3\arcsec \times 3\arcsec$ VISIR images of type I AGN 
             that do not show extra-nuclear emission. Each row shows one 
             object, sorted by Right Ascension. From left to right, the 
             images are sorted by filter central wavelength.}
    \label{sy1imas}
  \end{center}
\end{figure*}

\begin{figure*}
  \begin{center}
    \includegraphics[height=24cm]{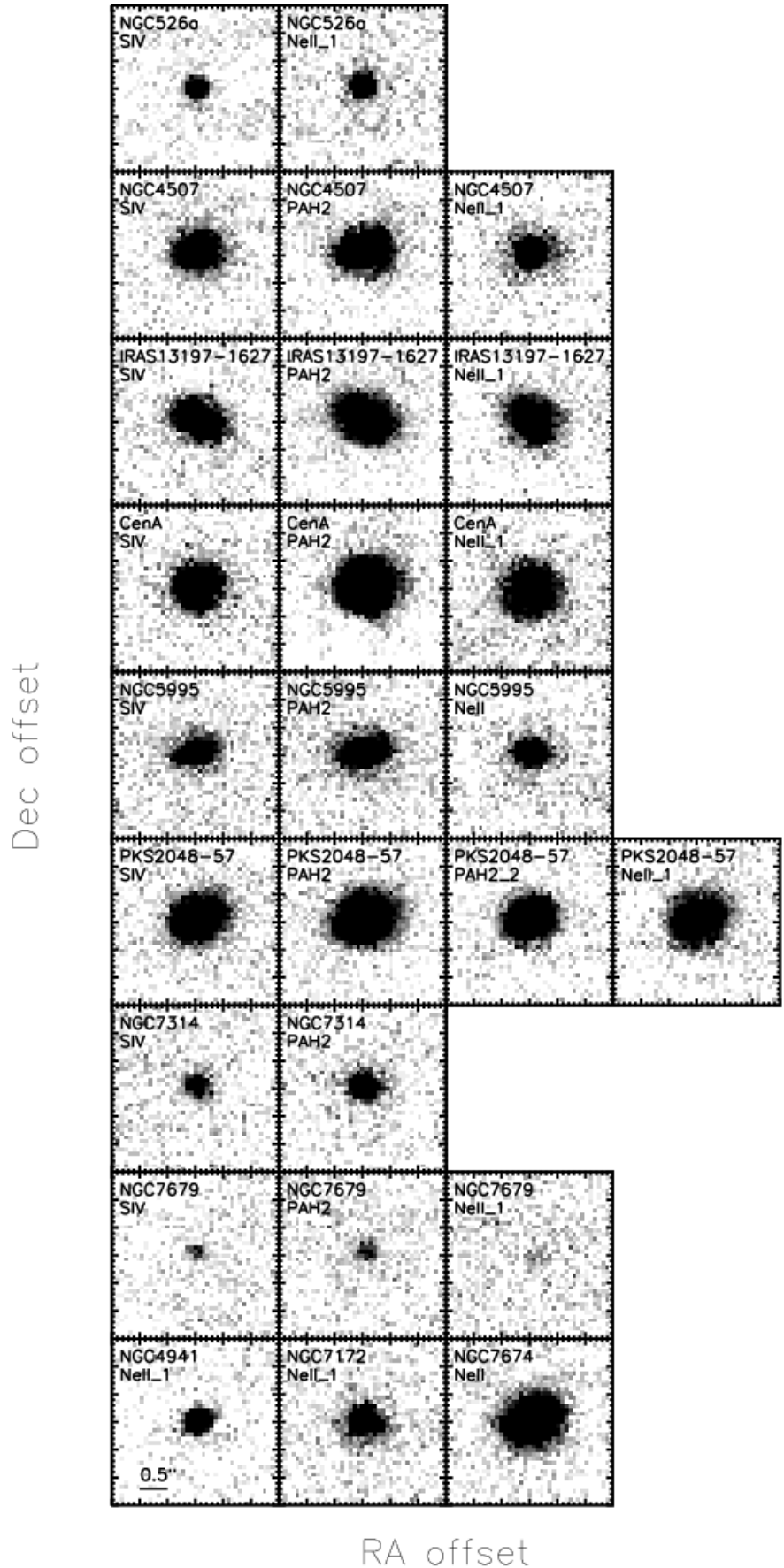}
    \caption[VISIR images of type II AGN]
            {$3\arcsec \times 3\arcsec$ VISIR images of type II AGN 
             that do not show extra-nuclear emission. Each row shows one 
             object, sorted by Right Ascension. From left to right, the 
             images are sorted by filter central wavelength. An exception to
             this is the last row which shows the three objects that were only
             observed in one filter.}
    \label{sy2imas}
  \end{center}
\end{figure*}

\begin{figure*}
  \begin{center}
    \includegraphics[width=10cm]{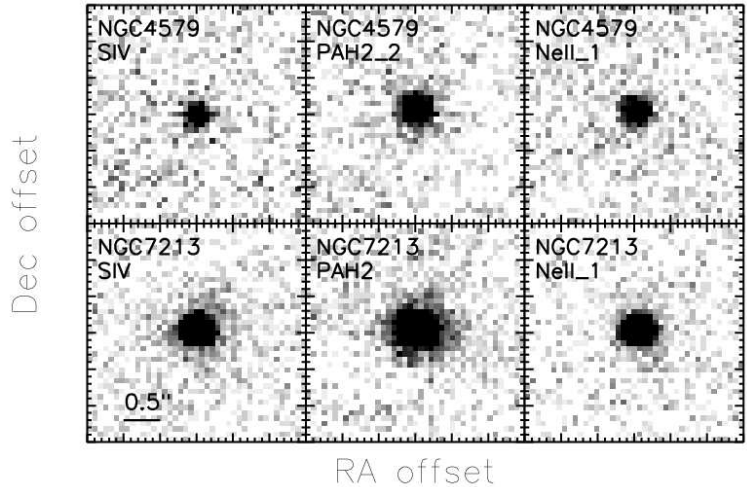}
    \caption[VISIR images of LINER AGN]
            {$3\arcsec \times 3\arcsec$ VISIR images of LINER type AGN that do
             not show extra-nuclear emission. Each row shows one 
             object, sorted by Right Ascension. From left to right, the 
             images are sorted by filter central wavelength.}
    \label{linerimas}
  \end{center}
\end{figure*}

\begin{figure}
  \begin{center}
    \includegraphics[width=0.4\textwidth]{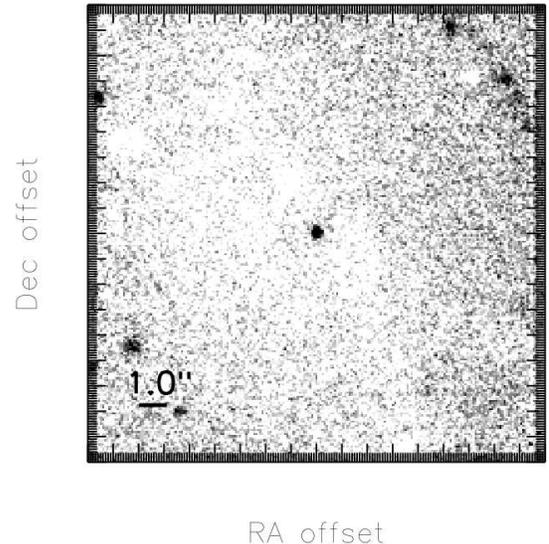}
    \caption[VISIR image of NGC 1097]
            {$16\farcs5 \times 16\farcs5$ VISIR image of NGC 1097, taken in
            the NeIIref1 filter.}
    \label{ngc1097ima}
  \end{center}
\end{figure}

\begin{figure}
  \begin{center}
    \includegraphics[width=0.4\textwidth]{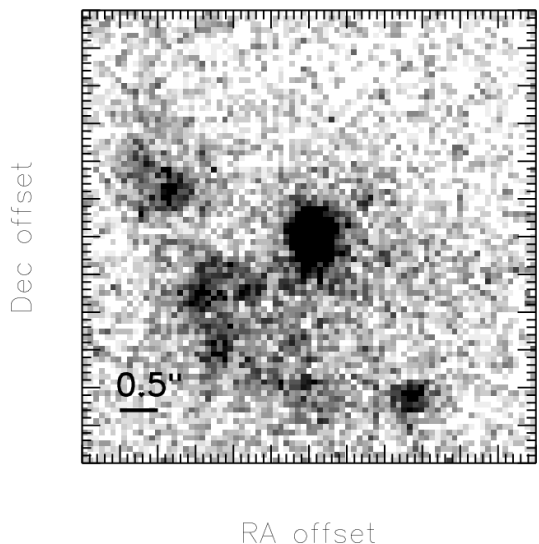}
    \caption[VISIR image of NGC 5135]
            {$6\arcsec \times 6 \arcsec$ VISIR image of NGC 5135, taken in the
             NeIIref1 filter.}
    \label{ngc5135ima}
  \end{center}
\end{figure}

\begin{figure*}
  \begin{center}
    \includegraphics[width=10cm]{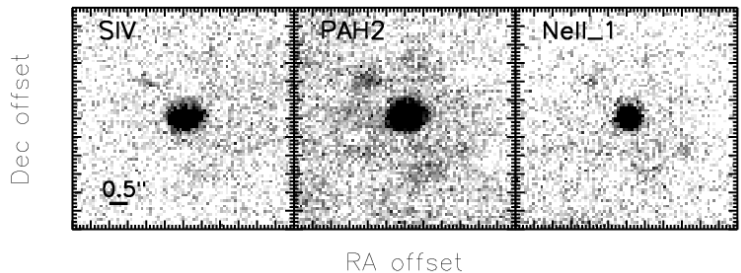}
    \caption[VISIR images of NGC 7469]
            {$6\arcsec \times 6 \arcsec$ VISIR images of NGC 7469, taken in the
             SIV, PAH2 and NeIIref1 filters.}
    \label{ngc7469ima}
  \end{center}
\end{figure*}

All recorded images are displayed in Figs. \ref{sy1imas} to \ref{ngc7469ima}. 
First, we show the type I, type II and LINER AGN that do not exhibit 
extranuclear emission (Figs. \ref{sy1imas}, \ref{sy2imas} and \ref{linerimas}, 
respectively), then we show the three objects that do exhibit extranuclear
emission, namely \object{NGC 1097} (Fig. \ref{ngc1097ima}), \object{NGC 5135}
(Fig. \ref{ngc5135ima}) and \object{NGC 7469} (Fig. \ref{ngc7469ima}). The 
images are linearly scaled. The minimum (white colour) is set to the mean 
background value $\left< \mathrm{BG} \right>$. The maximum (black colour) is 
set to $\left< \mathrm{BG} \right> + 5 \, \sigma_{\mathrm{BG}}$ where 
$\sigma_{\mathrm{BG}}$ is the standard deviation of the background.

Most objects appear point-like and the torus remains unresolved in all
cases. Refer to Table 2 in \citetalias{horst06} and Table 3 in
\citetalias{horst08a} for a comparison of object and standard star PSFs. 

In some cases the observing conditions were good enough to provide us with
diffraction limited imaging. In the images of NGC 3783 and NGC 4593 (third 
and fourth row, respectively, in Fig. \ref{sy1imas}), and NGC 4507 (second row
in Fig. \ref{sy2imas}), we can see hints of the first Airy ring. It is also 
barely visible in a number of other images. Slight elongation of the central 
source is visible in \object{IRAS 13197-1627} and \object{NGC 5995} (third and 
fifth row, respectively, in  Fig. \ref{sy2imas}). Extranuclear emission is 
observed in \object{NGC 1097} (Fig. \ref{ngc1097ima}), \object{NGC 5135} (Fig. 
\ref{ngc5135ima}) and \object{NGC 7469} (Fig. \ref{ngc7469ima}). These objects 
will be discussed in more detail in subsection \ref{mirmorph}. 

The slight elongation seen in the images of \object{NGC 5995} and \object{IRAS
13197-1627} is noteworthy. However, similar deformations of the PSF have been
seen with VISIR even for standard stars. The reason for this effect is not 
known; one possibility is tilt anisoplanatism as described by 
\citet{tokovinin07}. Since the instability of the PSF is a known problem with
VISIR, we suspect that the elongation is caused by instrumental effects in
both cases. These 2 objects do not belong to the ``well resolved sources''
sample defined in \citetalias{horst08a}.

The spectral properties of the observed objects will be discussed in section
\ref{spitzer}. Please note that we only chose the NeII filter for objects 
which are sufficiently redshifted. Thus, our photometry is not affected by
[Ne\scriptsize II\normalsize ] line emission.

\subsection{Morphologies of extranuclear emission} \label{mirmorph}

Three of the sources in our sample show extended extra-nuclear emission:
\object{NGC 1097}, \object{NGC 5135} and \object{NGC 7469}. In all three 
cases, we see distinct knots of star formation (SF) around the nucleus. 
Typical distances between SF knots and the AGN are $8.2\arcsec$ ($\sim 700$
pc) in \object{NGC 1097}, $1.5\arcsec$ ($\sim 400$ pc) in \object{NGC 5135} 
and $1.3\arcsec$ ($\sim 400$ pc) in \object{NGC 7469}. In the latter two 
objects the proximity of the SF components to the AGN as well as the presence 
of a weak diffuse component does not allow us to rule out a significant 
contribution of SF to the measured flux of the central point source. We have 
estimated this contribution by assuming that the flux of the SF component on 
top of the AGN does not exceed the flux of the brightest distinct SF knot. In 
the NeIIref1 filter, we find that the contamination is at most 15 \% for 
\object{NGC 5135} and 10 \% for \object{NGC 7469} within the aperture used for
the photometry. 
 
The SF we observe in \object{NGC 1097} (Fig. \ref{ngc1097ima}) is part of the 
well-known starburst ring of this galaxy. Unfortunately, when used with the
small field objective, the VISIR field-of-view is too small to image the whole
ring. Moreover, as the chop throw is of the same order of magnitude as the
separation between the AGN and the SF regions, positive and negative images of
different structures become intermingled. Therefore, we are not able to study
the MIR properties of the starburst in \object{NGC 1097}.

\begin{figure}
  \begin{center}
    \includegraphics[width=0.5\textwidth]{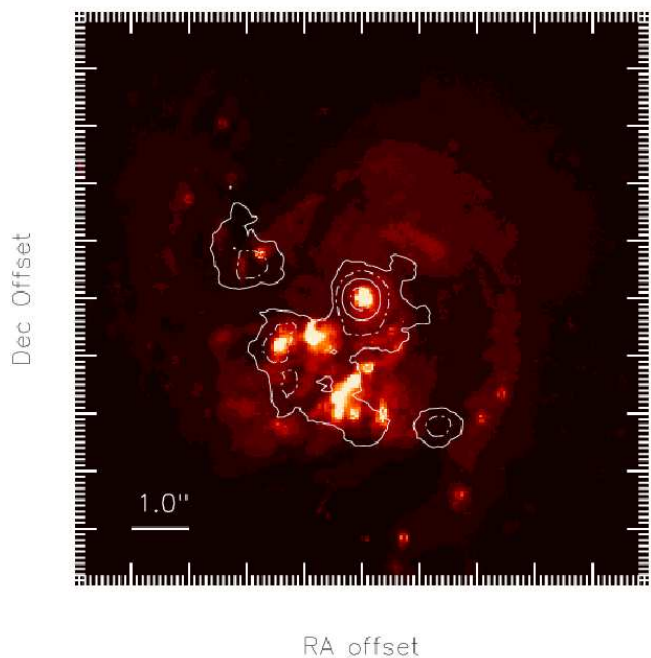}
    \caption[Overlay of HST and VISIR images of NGC 5135]{Overlay graphics,
    showing an HST image at $606$ nm of NGC 5135 with over-plotted VISIR
    contours (NeIIref1 filter). The contour levels are displayed as alternating
    solid and dashed-dotted lines. The image shows the central $10\arcsec 
    \times 10 \arcsec$, North is up and East is left. The VISIR contours only 
    cover the central $6\arcsec \times 6\arcsec$ of the displayed region.}
    \label{ngc5135overlay}
  \end{center}
\end{figure}

In \object{NGC 5135}, we see an arc of star formation at a distance of 
$\sim 400$ pc from the AGN. In Fig. \ref{ngc5135overlay}, we show an overlay of
optical and our MIR data. The optical image was taken with the WFPC2
instrument aboard the \emph{Hubble} Space Telescope (HST), using the F606W
filter. The MIR contours correspond to the NeIIref1 image, smoothed over 5
pixels. Alternating solid and dashed-dotted lines have been used for the
contours in order to avoid confusion between different morphological
features. For the relative astrometry, we matched the VISIR point source
to the closest point-like source on the Hubble image. Assuming that the VISIR
source is visible in the optical at all, the identification was
unambiguous. Possible offsets of the mid-IR and optical, expected if
torus and accretion disc are not aligned, are well below the resolution limit 
of both images. 

The HST image shows the spiral structure of the host galaxy, with
the AGN residing in the central bar. To the South, we find a very active SF
region. Here, optical and mid-infrared morphologies are very similar. Two
other MIR sources -- one to the East and one to the Southwest of the nucleus,
however, have only weak or no optical counterparts. In both cases, we likely
see SF that is embedded in thick layers of dust. The dust absorbs the emission
of the hidden young stars and thermally re-emits in the MIR regime. It is 
interesting to note that the region of SF is neither aligned to the spiral 
arms of the galaxy, nor does it show the characteristic ring-like structure 
often found around AGN. 

\begin{figure}
  \begin{center}
    \includegraphics[width=0.5\textwidth]{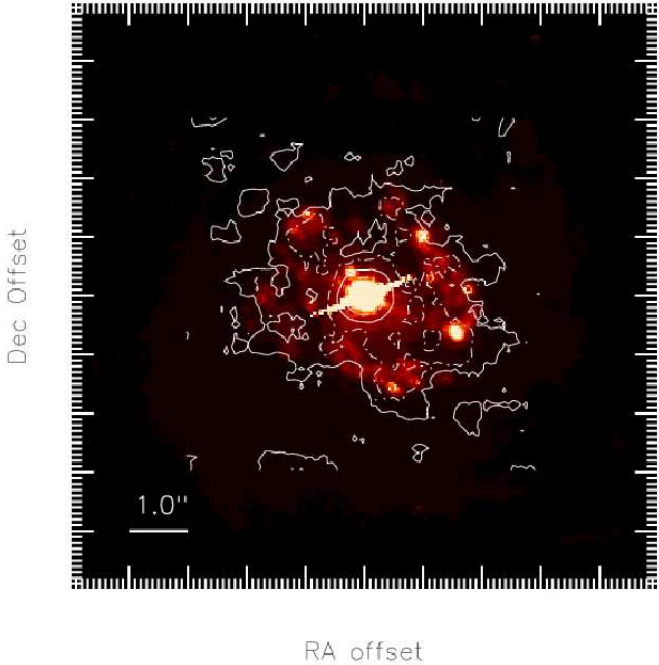}
    \caption[Overlay of HST and VISIR images of NGC 7469]{Overlay graphics,
    showing an HST image at $606$ nm of NGC 7469 with over-plotted VISIR
    contours (PAH2 filter).  The contour levels are displayed as alternating 
    solid and dashed-dotted lines. The image shows the central $10\arcsec 
    \times 10 \arcsec$, North is up and East is left. The VISIR contours only 
    cover the central $6\arcsec \times 6\arcsec$ of the displayed region.}
    \label{ngc7469overlay}
  \end{center}
\end{figure}

Such a ring of SF is found in \object{NGC 7469} as can be seen in Fig. 
\ref{ngc7469overlay}, where we show an overlay of an optical HST image and a 
VISIR NeIIref1 image. This overlay has been made in the same way as the one 
for NGC 5135. Again, we only find a partial coincidence of optical and MIR 
morphologies. Some of the less bright optical knots are strong MIR emitters, 
indicating SF covered by dust. It should be noted, however, that the S/N of
the VISIR image in the starburst ring is rather low, not allowing for a 
detailed matching of morphological features.

For the two cases of \object{NGC 5135} and \object{NGC 7469}, we find that 
within $3 \arcsec$ from the nucleus, SF contributes at least 43 \% and 45 \% 
of the total continuum flux in the NeIIref1 filter, respectively. These 
numbers can only be lower limits due to the observational limitations 
discussed above.

\subsection{Comparison to \emph{Spitzer} data} \label{spitzer}

We have browsed the \emph{Spitzer} science archive for low (spectral) 
resolution spectra of our sources, taken with the \emph{Infrared Spectrograph} 
(IRS) aboard the \emph{Spitzer} Space Telescope. With its 85 cm mirror, the 
angular resolution of \emph{Spitzer} is about 10 times less than the one of 
the VLT. On the other hand, space-borne IR observatories are superior in terms 
of sensitivity and spectral coverage.

The IRS spectra allow us to look for mid-infrared emission lines and 
compare the spectral shapes and flux levels of our data to the spatially less
resolved \emph{Spitzer} observations. To this end, we downloaded reduced 
data and extracted the spectra with SPICE 2.0.4, using the ``point source 
with regular extract'' generic template. The absolute flux calibration was done
by matching the automatic flux calibration SPICE provides for the overlapping
parts of different spectral settings. With this method, we achieved an 
absolute flux uncertainty of $\sim 50$ mJy. Especially large discrepancies 
between different spectral settings were found for \object{Cen A}, 
\object{MCG-06-30-15} and \object{NGC 3783}. For \object{NGC 4579} and 
\object{NGC 5995}, the \emph{Spitzer} science archive contains only
observations made in one spectral setting as of November 2007.

\begin{table*}
  \begin{center}
    \caption{Flux difference $F_{\mathrm{IRS}} - F_{\mathrm{VISIR}}$ in mJy and
             in units of the error on the VISIR flux measurements. Where the
             difference is significant, it is put in bold face. For these 
	     cases, in brackets we also state it in units of the IRS flux.}
    \scriptsize{
    \begin{tabular}{lcr|cr|cr|cr|cr|cr}
      \hline\hline
      Object & \multicolumn{2}{c}{SIVref1} & \multicolumn{2}{c}{SIV} &
        \multicolumn{2}{c}{PAH2} & \multicolumn{2}{c}{PAH2ref2} &
        \multicolumn{2}{c}{NeIIref1} & \multicolumn{2}{c}{NeII} \\
      & [mJy] & [$\sigma_{\mathrm{VISIR}}$] & [mJy] & 
        [$\sigma_{\mathrm{VISIR}}$] & [mJy] & [$\sigma_{\mathrm{VISIR}}$] &
        [mJy] & [$\sigma_{\mathrm{VISIR}}$] & [mJy] & 
        [$\sigma_{\mathrm{VISIR}}$] & [mJy] & [$\sigma_{\mathrm{VISIR}}$]  \\
      \hline
      Fairall 9 & \multicolumn{2}{c|}{--} & 5.58 & 3.16 & 
        \multicolumn{2}{c|}{--} & \multicolumn{2}{c|}{--} & 29.65 & 2.86 &
        38.44 & 2.95 \\
      NGC 526a & \multicolumn{2}{c|}{--} & -29.19 & 1.14 & 
        \multicolumn{2}{c|}{--} & \multicolumn{2}{c|}{--} & -47.20 & -0.86  &
        \multicolumn{2}{c}{--} \\
      {\bfseries Mrk 590} & \multicolumn{2}{c|}{--} & 60.98 & 2.92 & 
        \multicolumn{2}{c|}{--} & {\bfseries 80.81} (0.52)& {\bfseries 38.53} &
        \multicolumn{2}{c|}{--} & {\bfseries 67.56} (0.61)& {\bfseries 5.20} \\
      NGC 3783 & \multicolumn{2}{c|}{--} & -37.61 & 0.81 & 
        \multicolumn{2}{c|}{--} & 15.28 & 0.70 & -47.38 & 0.70 & 
        \multicolumn{2}{c}{--} \\
      NGC 4507 & \multicolumn{2}{c|}{--} & -44.11 & 1.25 & -52.00 & 1.69 &
        \multicolumn{2}{c|}{--} & -60.85 & 1.31 & \multicolumn{2}{c}{--} \\
      {\bfseries NGC 4579} & \multicolumn{2}{c|}{--} & 31.84 & 1.84 & 
        \multicolumn{2}{c|}{--} & {\bfseries 51.34} (0.43)& {\bfseries 3.71} & 
        {\bfseries 66.09} (0.52)& {\bfseries 3.21} & \multicolumn{2}{c}{--} \\
      NGC 4593 & \multicolumn{2}{c|}{--} & 31.71 & 1.10 & 
        \multicolumn{2}{c|}{--} & 70.20 & 2.69 & 29.58 & 0.40 &
        \multicolumn{2}{c}{--} \\
      IRAS 13197-1627 & \multicolumn{2}{c|}{--} & 28.19 & 1.65 & 43.39 & 1.21 &
        \multicolumn{2}{c|}{--} & -9.82 & 0.21 & \multicolumn{2}{c}{--} \\
      {\bfseries Cen A} & \multicolumn{2}{c|}{--} & {\bfseries 193.77} (0.23)& 
        {\bfseries 7.28} & {\bfseries 418.28} (0.30)& {\bfseries14.32} & 
        \multicolumn{2}{c|}{--} & {\bfseries 421.32} (0.23)& {\bfseries 5.76} &
        \multicolumn{2}{c}{--} \\
      {\bfseries MCG-06-30-15} & \multicolumn{2}{c|}{--} & {\bfseries 61.00} (0.85)&
        {\bfseries 4.42} & 35.04 & 1.27 & \multicolumn{2}{c|}{--} & 43.19 & 
        0.95 & \multicolumn{2}{c}{--} \\
      NGC 5995 & \multicolumn{2}{c|}{--} & -37.18 & 1.23 & -10.03 & 0.21 & 
        \multicolumn{2}{c|}{--} & \multicolumn{2}{c|}{--} & -38.59 & 0.64 \\
      {\bfseries Mrk 509} & \multicolumn{2}{c|}{--} & 34.36 & 4.48 & 
        {\bfseries 69.53} (0.23)& {\bfseries 3.25} & \multicolumn{2}{c|}{--} & 
        \multicolumn{2}{c|}{--} & 86.67 & 2.08\\
      PKS 2048-57 & \multicolumn{2}{c|}{--} & -56.27 & 2.04 & -15.72 & 0.35 &
        -32.17 & 0.61 & -118.73 & 1.86 & \multicolumn{2}{c}{--} \\
      PG 2130+099 & 11.88 & 0.49 & \multicolumn{2}{c|}{--} & 3.80 & 0.48 & 
        \multicolumn{2}{c|}{--} & \multicolumn{2}{c|}{--} & 9.49 & 0.31 \\
      NGC 7213 & \multicolumn{2}{c|}{--} & 4.92 & 0.79 & 56.49 & 1.47 & 
        \multicolumn{2}{c|}{--} & 29.29 & 1.11 & \multicolumn{2}{c}{--} \\
      3C 445 & \multicolumn{2}{c|}{--} & -0.72 & 0.10 & 14.01 & 1.40 & 
        \multicolumn{2}{c|}{--} & \multicolumn{2}{c|}{--} & 21.60 & 0.77 \\
      NGC 7314 & \multicolumn{2}{c|}{--} & 2.69 & 0.07 & 28.38 & 1.29 & 
        \multicolumn{2}{c|}{--} & \multicolumn{2}{c|}{--} & 
        \multicolumn{2}{c}{--} \\
      {\bfseries NGC 7469} & \multicolumn{2}{c|}{--} & {\bfseries 133.69} (0.23)& 
        {\bfseries 6.68} & {\bfseries 430.58} (0.47)& {\bfseries 11.15} & 
        \multicolumn{2}{c|}{--} & {\bfseries 373.48} (0.37)& {\bfseries 10.76} &
        \multicolumn{2}{c}{--}\\
      {\bfseries NGC 7679} & \multicolumn{2}{c|}{--} & {\bfseries 71.89} (0.63)& 
        {\bfseries 5.53} & {\bfseries 212.38} (0.83)& {\bfseries 32.18} & 
        \multicolumn{2}{c|}{--} & {\bfseries 193.70} (0.81)& {\bfseries 10.58} &
        \multicolumn{2}{c}{--} \\
      \hline
    \end{tabular}}
    \label{offsets}
  \end{center}
\end{table*}

\begin{figure*}
  \begin{center}
    \includegraphics[width=\textwidth]{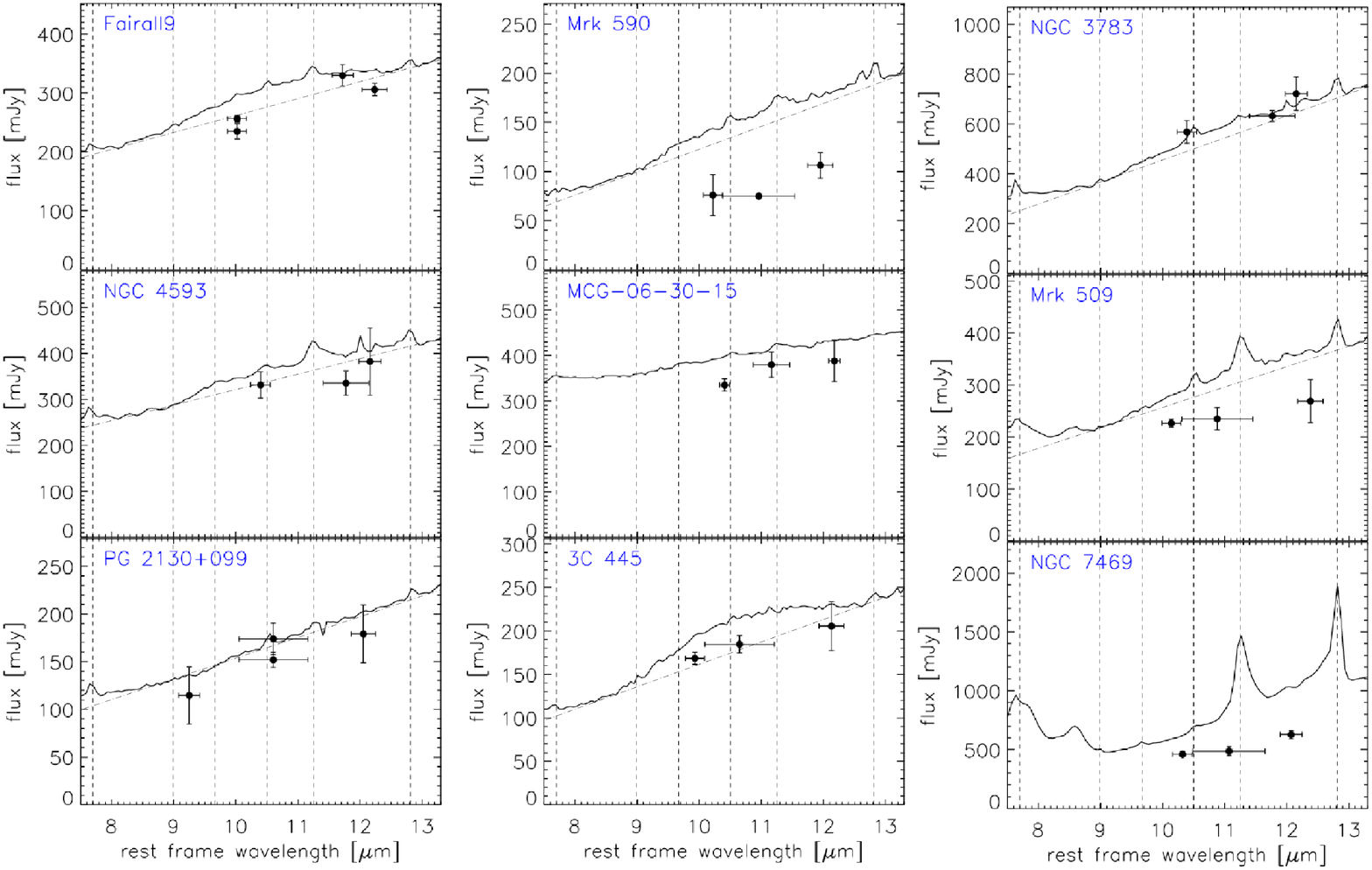}
    \caption[\emph{Spitzer} spectra of type I AGN]{Comparison of low
             resolution \emph{Spitzer} IRS spectra (solid lines) and our VISIR 
             photometry (filled circles) for the type I AGN among our sample 
             for which both data sets are available. Wavelengths are in rest 
             frame, fluxes as observed. Horizontal error bars
             correspond to the filter pass band. The absolute flux uncertainty 
             of the IRS spectra is about 50 mJy. Dashed-dotted lines indicate
             the level of the continuum. Vertical dashed lines denote the 
             position of common emission lines: $7.7 \, \mu$m PAH, 
             $8.99 \, \mu$m [Ar\scriptsize III\normalsize], $9.67 \, \mu$m 
             H$_{2}$ (0-0)S3, $10.5 \, \mu$m [S\scriptsize IV\normalsize], 
             $11.3 \, \mu$m PAH and $12.81 \, \mu$m [Ne\scriptsize 
             II\normalsize], in some cases blended with $12.7 \, \mu$m PAH.}
    \label{spitzer_sy1}
  \end{center}
\end{figure*}

\begin{figure*}
  \begin{center}
    \includegraphics[width=\textwidth]{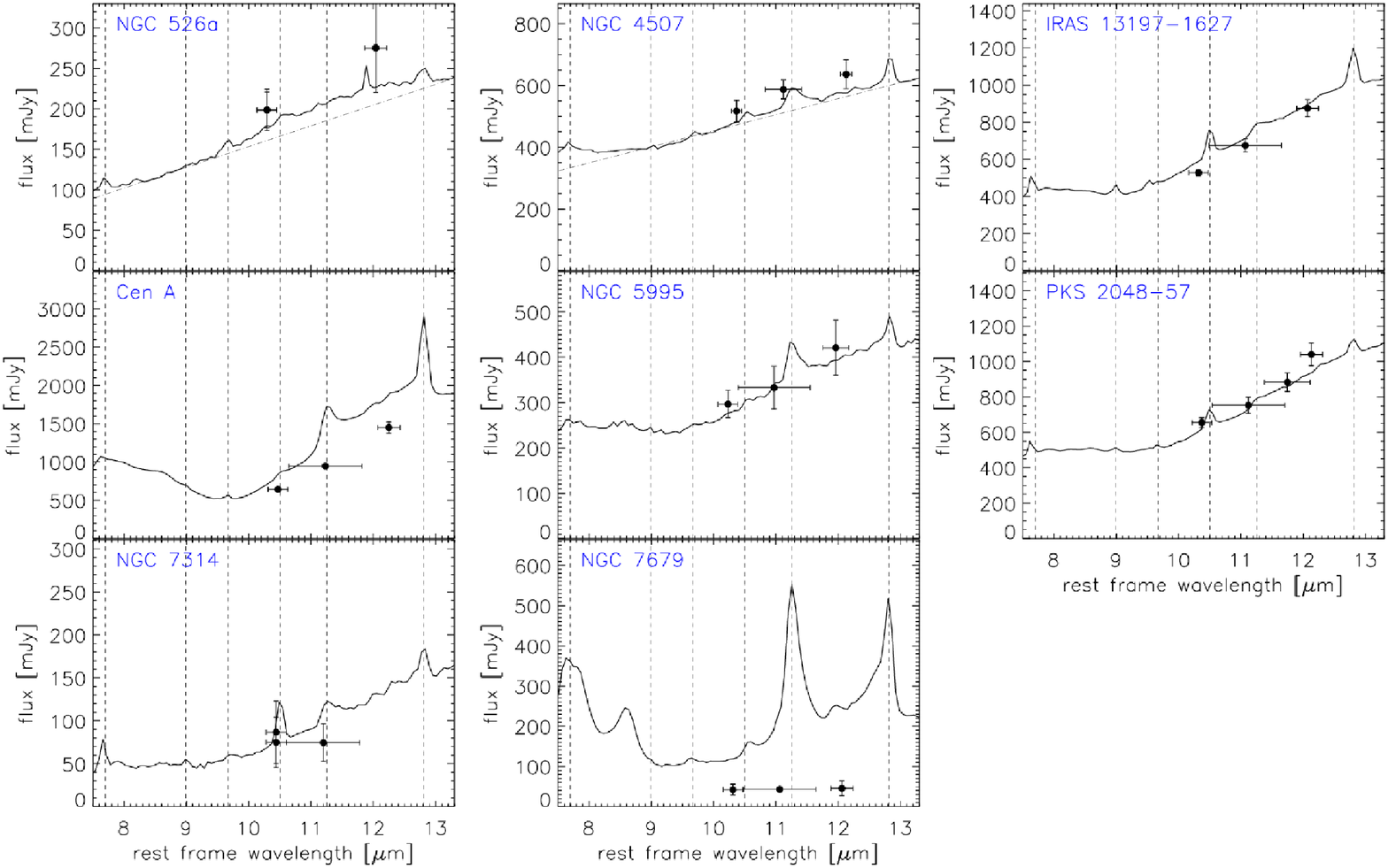}  
    \caption[\emph{Spitzer} spectra of type II AGN]{Comparison of low
             resolution \emph{Spitzer} IRS spectra (solid lines) and our VISIR 
             photometry (filled circles) for the type II AGN among our sample 
             for which both data sets are available. Wavelengths are in rest 
             frame, fluxes as observed. Horizontal error bars
             correspond to the filter pass band. The absolute flux uncertainty 
             of the IRS spectra is about 50 mJy. Dashed-dotted lines indicate
             the level of the continuum. Vertical dashed lines denote the
             location of the same emission lines as in fig. \ref{spitzer_sy1}.}
    \label{spitzer_sy2}
  \end{center}
\end{figure*}

\begin{figure}
  \begin{center}
    \includegraphics[width = 0.5\textwidth]{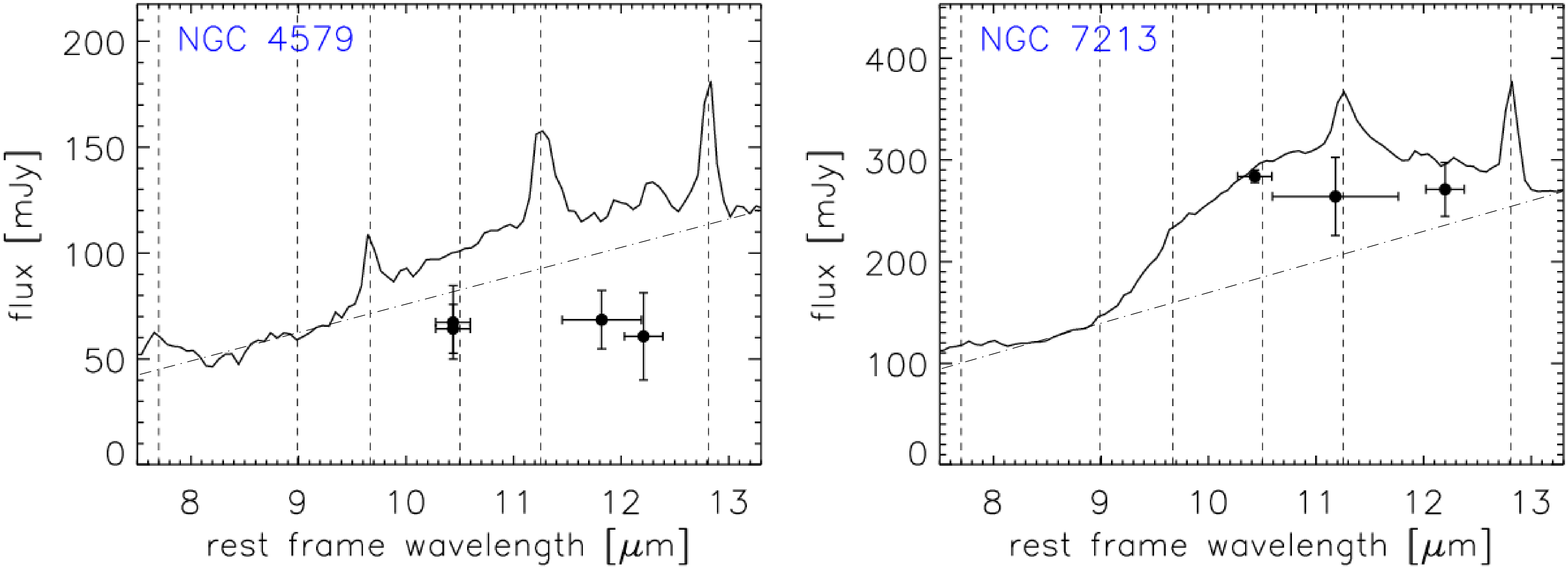}
    \caption[\emph{Spitzer} spectra of LINER AGN]{Comparison of low
             resolution \emph{Spitzer} IRS spectra (solid lines) and our VISIR 
             photometry (filled circles) for the LINER AGN among our sample for
             which both data sets are available. Wavelengths are in rest 
             frame, fluxes as observed. Horizontal error bars
             correspond to the filter pass band. The absolute flux uncertainty 
             of the IRS spectra is about 50 mJy. Dashed-dotted lines indicate
             the level of the continuum. Vertical dashed lines denote the
             location of the same emission lines as in fig. \ref{spitzer_sy1}.}
    \label{spitzer_liners}
  \end{center}
\end{figure}

In figs. \ref{spitzer_sy1}, \ref{spitzer_sy2} and \ref{spitzer_liners}, we
show IRS spectra and our VISIR photometry for all objects which we had
observed in two or more different filters and for which IRS spectra were 
publicly available. We compare \emph{Spitzer} and VISIR data of 9 type I AGN
(Fig. \ref{spitzer_sy1}), 8 type II AGN (Fig. \ref{spitzer_sy2}) and 2 
LINERs (Fig. \ref{spitzer_liners}). The wavelength scale in all panels is in 
rest-frame, vertical dashed lines depict the positions of common mid-infrared 
emission lines: $7.7 \, \mu$m PAH, $8.99 \, \mu$m [Ar\scriptsize 
III\normalsize], $9.67 \, \mu$m H$_{2}$ (0-0)S3, $10.5 \, \mu$m [S\scriptsize 
IV\normalsize], $11.3 \, \mu$m PAH and $12.81 \, \mu$m [Ne\scriptsize 
II\normalsize], in some cases blended with $12.7 \, \mu$m PAH. In order to
increase the visibility of the silicate emission feature, we also plot a
linear interpolation (dashed-dotted line) with anchor points at $8.7 \, \mu$m
and $13.2 \, \mu$m for every object that might exhibit the feature (meant to 
roughly depict source continuum). In Table \ref{offsets}, we then show the 
difference between IRS and VISIR fluxes, both in absolute and relative units. 
To compare both flux measurements, the IRS fluxes were integrated over the 
width of the according VISIR filter. We then find 7 out of 19 objects to 
exhibit a flux difference of more than $3 \, \sigma_{\mathrm{VISIR}}$ and more 
than 50 mJy, in at least one filter. In no case do we find a VISIR flux larger 
at more than $3 \sigma$ and more than 50 mJy than the \emph{Spitzer} IRS flux.

Among the type I AGN, we find a relatively good agreement between 
\emph{Spitzer} and VISIR data, despite the difference in resolution. The flux
level is in significant disagreement for \object{Mrk 590}, 
\object{MCG-06-30-15}, \object{Mrk 509} and \object{NGC 7469}. Presumably, in 
these cases the IRS spectra contain contribution from SF around the nucleus. 
In \object{NGC 7469}, \emph{Spitzer} is not able to resolve the SF ring we see 
with VISIR. The strong PAH and [Ne\scriptsize II\normalsize] emission lines in 
this object indicate the strong contribution of SF to the measured flux. 
Moreover, the flux discrepancy found is in good agreement with SF flux we 
estimated from our VISIR data in section \ref{mirmorph}. In no case do we find 
a VISIR flux larger than the \emph{Spitzer} one. In a number of other objects, 
\emph{Spitzer} reveals the presence of H$_2$, PAH and 
[Ne\scriptsize II\normalsize] emission lines. In addition, the $9.7 \, \mu$m  
silicate emission feature is present in the spectra of \object{Fairall 9},
\object{Mrk 590}, \object{NGC 3783},\object{NGC 4593}, \object{Mrk 509} and
\object{3C 445}. As has been observed earlier by
\citet{haol05,siebenmorgen05} and \citet{sturm05}, the centre of the feature
is shifted redward to $\sim 10.5 \, \mu$m. 

For our type II sources, the IRS spectra show emission lines in every
object. Especially remarkable is the spectrum of NGC 7679 which resembles the
one of a starburst galaxy rather than an AGN. The VISIR photometry, however,
does not appear to be significantly affected by SF as we do not find any
indication for the presence of the $11.3 \, \mu$m PAH line which is very
strong in the IRS spectrum. The $9.7 \, \mu$m silicate absorption feature is
present in the spectra of \object{IRAS 13197-1627}, \object{Cen A},
\object{NGC 5995}, \object{PKS 2048-57} and \object{NGC 7314}. Interestingly, 
\object{NGC 526a} may show the feature in emission rather than absorption. It 
is absent in \object{NGC 4507} and the starburst-like spectrum of 
\object{NGC 7679}. Please note that in case of \object{NGC 4507}, a 
re-examination of the \emph{Spitzer} data set has shown that this observation 
was slightly mis-pointed (M. Haas, private communication). Thus, the actual 
nuclear spectral shape might be different than shown in Fig. \ref{spitzer_sy2}.
This would also explain why the VISIR fluxes are slightly larger than the IRS 
ones. It seems possible that the same has happened for \object{PKS 2048-57}
and \object{NGC 526a} for which the fluxes measured with VISIR are slightly -- 
but not significantly -- larger than the ones observed with IRS, as well.  

The two LINERs in our sample appear quite differently in VISIR and
\emph{Spitzer} observations. The IRS spectrum of \object{NGC 4579} shows a 
rising continuum toward the red and a silicate emission feature while the 
VISIR measurements show a flat continuum. The strong 
[Ar\scriptsize III\normalsize], PAH and [Ne\scriptsize II\normalsize] lines 
indicate the presence of active SF. Cold dust in the outer parts of a deeply 
embedded SF region could also explain the red continuum colour. In the case of 
\object{NGC 7213}, the continuum levels measured with VISIR and \emph{Spitzer} 
are in agreement with each other. In the IRS spectrum, a silicate emission 
feature is superimposed onto the strong rise toward longer wavelengths. This 
feature appears to be absent in the VISIR photometry. 

While a quantitative analysis of the difference in flux between IRS spectra and
VISIR photometry is difficult due to the lacking accuracy of the absolute flux
calibration we have performed on the IRS data, we do find that in at least 7 
cases, the disagreement is significant. In these cases, between 
$(0.20 \sim 0.85) \cdot F_{\mathrm{IRS}}$ cannot be accounted for with VISIR 
and, thus, very likely originates in circumnuclear SF or other 
extra-nuclear phenomena. Interestingly, we do not find a clear dependence of 
the discrepancy between VISIR and IRS fluxes on the scale we resolve with 
VISIR, in terms of the dust sublimation radius.

The cases of \object{Mrk 590} and \object{MCG-06-30-15} are of particular 
interest in this respect as the IRS spectra of these objects do not exhibit 
strong emission lines indicative of star formation (i.e. PAH, 
[Ne\scriptsize II\normalsize]). This means that the absence of such lines does 
not necessarily imply a spectrum to be free of contaminating, extra-nuclear 
emission.

\section{Discussion} \label{discussion}

\subsection{Extended emission} \label{extem}

The strength of the extended MIR emission in \object{NGC 5135} (Fig. 
\ref{ngc5135ima}) and \object{NGC 7469} (Fig. \ref{ngc7469ima}) illustrates
the importance of high angular resolution for AGN studies that has been
discussed in \citetalias{horst08a}. This point is also demonstrated by the
difference in flux level we see between VISIR and \emph{Spitzer} observations
of some of our targets. Moreover, \object{Mrk 590} and \object{MCG-06-30-15} 
indicate that significant contamination by circumnuclear emission can be 
present in MIR data even when IR emission lines are weak. Therefore, satellite 
IR data of AGN should be treated with care when directly compared to torus 
models.

\subsection{The silicate emission feature in Seyfert AGN} \label{silicates}

As we have seen in section \ref{spitzer}, 6 out of 9 type I AGN displayed in
Fig. \ref{spitzer_sy1} show the $9.7 \, \mu$m silicate feature in emission. 
Thus, silicate emission seems to be quite common in type I AGN of moderate 
luminosities, contrary to suggestions they may only be pronounced in QSOs 
\citep{siebenmorgen05}. Furthermore, within our limited sample statistics, we
do not find a clear trend of more luminous AGN to exhibit stronger silicate
features. Although the most luminous object shown in Fig. 9 -- \object{3C 445}
-- also exhibits the strongest silicate emission feature, there seems to be no
trend for the rest of the sample. Also note that \object{PG 2130+099} does not 
show the feature at all while the less luminous objects \object{Mrk 590},
\object{NGC 3783} and \object{NGC 4593} do \citepalias[see][for the
luminosities]{horst08a}. 

While we find silicate emission to be common among type I Seyferts, the
features are generally weak. Comparing this result to radiative transfer 
calculations of different torus models indicates that the dust is
distributed in clumps rather than smoothly 
\citep[see e.g.][]{nenkova02,hoenig06}.

In addition to type I AGN, we see the silicate emission feature in the LINERs 
\object{NGC 4579} and \object{NGC 7213}. This is another indication that the 
strength of the feature is not correlated to AGN luminosity. Most remarkable 
is the possible presence of a silicate emission feature in the type II AGN 
\object{NGC 526a}. This adds to the observations of silicate emission in 
luminous type II QSOs by \citet{sturm06} and \citet{teplitz06}. Smooth torus 
models, on the other hand, predict type II AGN to only show the feature in 
absorption and not in emission.

In \citetalias{horst08a}, we discussed the relevance of high angular
resolution when AGN torus models are compared to mid-infrared
observations. The comparison between our VISIR data and archival 
\emph{Spitzer} data reveals another problem with low angular resolution 
studies of AGN: For some objects -- \object{NGC 4579} and \object{NGC 7213} 
 -- the strength of silicate feature strongly depends on the resolved scale. 

Recently, \citet{schweitzer08} discussed NLR dust clouds as the possible main
source of the silicate emission feature in PG QSOs. They find that the average 
distance $r_{\mathrm{dust}}$ of these dust clouds to the AGN is 
$r_{\mathrm{dust}} \sim 170 \, r_{\mathrm{sub}}$ where $r_{\mathrm{sub}}$ is 
the dust sublimation radius. This effect can explain both the presence of 
silicate emission in type II AGN such as \object{NGC 526a} and the difference 
in silicate feature strength between IRS and VISIR data. \object{NGC 4579} and
\object{NGC 7213} have PSF sizes of $\sim 2000 \, r_{\mathrm{sub}}$ and 
$\sim 600 \, r_{\mathrm{sub}}$, respectively. Nevertheless, this could mean 
that for objects with very extended cloud distributions, we start to resolve 
these with VISIR and thus see a difference in feature strength between VISIR 
and \emph{Spitzer},  while we fail to resolve the cloud distribution for 
objects with $r_{\mathrm{dust}} = 170 \, r_{\mathrm{sub}}$. 

Neither the results of \citet{schweitzer08} nor our own rule 
out additional silicate emission from the torus itself. As the emission
features seen in Fig. \ref{spitzer_sy1} are weak to begin with, however, the
torus emission has to be low, especially when compared to prominent silicate
absorption features as in e.g. \object{Cen A} or \object{NGC 5995}. The
weakness of the feature is a strong argument in favour of clumpy torus models
since these predict a less pronounced feature than smooth ones
\citep[see][]{nenkova02,hoenig06}.

\subsection{Effect of the silicate feature on our continuum photometry}

In \citetalias{horst08a}, we claimed that our measured $12.3 \, \mu$m fluxes
were not affected by the silicate feature. Despite its presence in many
objects, this claim still holds. In the VISIR photometry of all type I
sources, the feature is either absent or weak -- even in the latter case, at
$12.3 \, \mu$m the possible contamination is below 10\%. For the type II
sources, the error induced by the silicate feature is slightly larger and may
may amount to roughly 10\% in \object{IRAS 13197-1627} and 
\object{Cen A}. The only object for which we may have induced a larger error 
than that is \object{NGC 7314}. This will still not significantly affect the 
results derived in that work. 

\section{Conclusions} \label{conclusions}

We present MIR images of 25 local AGN, obtained with VISIR at the VLT with
$\sim 0\farcs35$ resolution. We found two of the objects to be slightly
extended -- most likely due to an instrumental effect -- and 
three to exhibit extra-nuclear emission in addition to the central point 
source. We identify this emission with regions of intensive star formation.

For 20 AGN, we were able to reconstruct their N-band SEDs; we find that about
half of the type I AGN exhibit a silicate emission feature at $\sim 10.5 \,
\mu$m. The relative weakness of the feature and the fact that it is not
present in all type I sources gives support to models of clumpy tori
\citep{nenkova02,dullemond05,hoenig06}. This matches our conclusions from
\citetalias{horst08a}. 

Furthermore, by comparing our data to \emph{Spitzer} IRS spectra, in three
objects we find indications for the silicate emission feature to originate in 
an extended region. This as well as the detection of silicate emission in the
type II AGN \object{NGC 526a} supports the results by \citet{schweitzer08}
that the bulk of silicate emission may be emitted from dust clouds in the NLR
rather than the obscuring torus. 

The comparison with \emph{Spitzer} spectra also reveals a significant offset
in flux level between VISIR and IRS data in at least 7 out of 19 objects. Here,
we find the contamination to be of the order of $(0.20 \sim 0.85) \cdot 
F_{\mathrm{IRS}}$. This
indicates the presence of extra-nuclear MIR emission on scales of several tens
to hundreds of parsecs and underlines the need for high angular resolution in 
MIR studies of AGN. The cases of \object{Mrk 590} and \object{MCG-06-30-15}
show that even objects without strong emission lines can suffer from 
contamination.

\begin{acknowledgements}
We thank Dr. Eric Pantin for kindly providing us with his VISIR pipeline. 
We are indebted to an anonymous referee who greatly helped to improve the 
manuscript.
H.H. acknowledges support from DFG through SFB 439.
P.G. is a Fellow of the Japan Society for the Promotion of Science (JSPS).
This research made use of the NASA/IPAC Extragalactic Database (NED) which 
is operated by the Jet Propulsion Laboratory, California Institute of 
Technology, under contract with the National Aeronautics and Space 
Administration.
We acknowledge the usage of the HyperLeda database 
(http://leda.univ-lyon1.fr).
\end{acknowledgements}

\bibliographystyle{aa}
\bibliography{mybiblio}

\end{document}